\documentclass[aps,prl,twocolumn,amssymb,showpacs]{revtex4}

\usepackage{amsmath}
\usepackage{amssymb}
\usepackage{amsthm}
\usepackage{pstricks,pst-node,pst-plot}
\newpsobject{showgrid}{psgrid}{subgriddiv=1, griddots=5, gridlabels=6pt}
\usepackage{graphicx}

\renewcommand{\(}{\left(}
\renewcommand{\)}{\right)}
\renewcommand{\[}{\left[}
\renewcommand{\]}{\right]}
\newcommand{\f}[2]{\frac{#1}{#2}}

\newcommand{\la}{\langle}
\newcommand{\ra}{\rangle}
\newcommand{\bra}[1]{\la #1 |}
\newcommand{\ket}[1]{| #1 \ra}
\newcommand{\beq}{\begin{equation}}
\newcommand{\eeq}{\end{equation}}
\newcommand{\nol}{\nonumber \\ }
\newcommand{\ph}{\phantom{\mu}}

\begin{document}

\title{Quantum Radiation Reaction and the Green's Function
Decomposition}

\author{Atsushi Higuchi$^1$ and Giles~D.~R.~Martin$^2$}

\affiliation{Department of Mathematics, University of York,
Heslington, York YO10 5DD, United Kingdom\\ email:
$^1$ah28@york.ac.uk, $^2$gdrm100@york.ac.uk}

\date{August 4, 2006}

\begin{abstract}
We analyze the change in position (the position shift)
of the wave packet of a
charged scalar particle due to radiation reaction in the
$\hbar \rightarrow 0$ limit of quantum electrodynamics.  In particular,
we re-express the formula previously obtained
for the position shift in terms of Green's functions for the
electromagnetic field, thus
clarifying the relation between the quantum and classical derivations of
the radiation-reaction force.
\end{abstract}

\pacs{03.65.-w,12.20.-m}

\maketitle

\section{Introduction}

The interaction of a particle with its own field produces the
phenomenon of radiation reaction. This phenomenon
results in an extra force on the particle, a self-force, which
changes the equations of motion. In classical electrodynamics in
Minkowski spacetime this extra force is known as the Lorentz-Dirac
force~\cite{Abraham,Lorentz,Dirac}.

Since classical electrodynamics is the classical limit of
quantum electrodynamics (QED), the Lorentz-Dirac force
should be derived from QED in the $\hbar \rightarrow 0$ limit.  This fact
was demonstrated in Refs.~\cite{Higuchi2,HM23,HM4} by showing that
the change in position (the position shift) due to
radiation reaction of a charged scalar particle in QED agrees with that
in classical electrodynamics, with the Lorentz-Dirac force treated as a
small perturbation,
in the $\hbar \rightarrow 0$ limit. (See
Refs.~\cite{MS,beilok,Tsyt,FordOc,Oc} for some
other approaches to find the
radiation-reaction force in QED.)
In this Letter we
analyze the expression for the quantum position shift obtained in
Refs.~\cite{HM23,HM4} to clarify the relation between the classical and
quantum theories of radiation reaction.
Our metric is
$g_{\mu\nu} = {\rm diag}\,(+1,-1,-1,-1)$, and we use Greek letters for
spacetime indices and Latin letters for space indices.

One of the methods for deriving the classical self-force
involves identification of the part of the self-field
responsible for the force~\cite{Dirac}.
(This method has been
extended to gravitational radiation~\cite{Whiting}.)
For classical electrodynamics in Minkowski spacetime,
the retarded Green's function $G_{-\mu\nu'}(x-x')$ satisfies the
equation
$\partial_\alpha\partial^\alpha G_{-\mu\nu'}(x-x') = \delta^4(x-x')$
and the
boundary condition $G_{-\mu\nu'}(x-x')=0$ for $x^0 < x^{\prime 0}$,
with primed indices corresponding to the primed spacetime coordinates.
The $G_{-}$ can be
decomposed into two parts, the regular and singular Green's functions,
$G_{\rm R}$ and $G_{\rm S}$:
\begin{equation}
G_{-\mu\nu'}(x-x') = G_{R\mu\nu'}(x-x') + G_{S\mu\nu'}(x-x')\,,
\end{equation}
with
$G_{{\rm R}}= \f{1}{2} \left( G_{-}- G_{+}\right)$ and
$G_{{\rm S}}= \f{1}{2} \left( G_{-} + G_{+}\right)$,
where $G_{+\mu\nu'}(x-x')$ is the advanced Green's function.
One shows that only
the field generated by the regular Green's function $G_{R\mu\nu'}(x-x')$
contributes to the radiation-reaction force, the field generated by the
singular Green's function leading to an infinite correction to
the inertial mass.  Thus, the effective
electromagnetic self-field contributing to the self-force is
\begin{align}\label{regularfield}
A_{{\rm R}\mu}(x) & = \int d^4 x'\,{G_{{\rm R}\mu}}^{\nu'}(x-x')
j_{\nu'}(x')\,,\\
j^\nu(x) & = e\frac{dx^\nu}{dt}\delta^3({\bf x}-{\bf X}(t))\,,
\label{current}
\end{align}
where ${\bf X}(t)$ is the position of the charged particle with charge
$e$ at time $t$. [Here we are using the notation
$x^\mu = (t,{\bf x})$.]
Let $F^{\mu\nu}_{\rm ext}$ be the external electromagnetic field that
accelerates the charged particle with mass $m$.
Then the equation
of motion for this particle modified by the Lorentz-Dirac force is
written as follows:
\begin{equation}
\frac{d\ }{dt}\left( m\frac{dX^\mu}{d\tau}\right)
= e F^{\mu\nu}_{\rm ext}\frac{dX^\mu}{dt}
+ f^\mu_{\rm LD}\,, \label{LorDir}
\end{equation}
where
\begin{equation}
f^\mu_{\rm LD} = eF_{\rm R}^{\mu\nu}\frac{dX_\nu}{dt}\,, \label{FLD}
\end{equation}
with $F_{{\rm R}\mu\nu} \equiv
\partial_\mu A_{{\rm R}\nu} - \partial_\nu A_{{\rm R}\mu}$.
Here $\tau$ denotes the proper-time.  Then it can be shown that
\begin{equation}
f^\mu_{\rm LD} = \frac{e^2}{6\pi}
\left[ \frac{d^3 X^\mu}{d\tau^3}
+ \frac{dX^\mu}{d\tau}
\left( \frac{d^2X^\nu}{d\tau^2}\frac{d^2X_\nu}{d\tau^2}\right)\right]
\frac{d\tau}{dt}\,. \label{FLD2}
\end{equation}
(See, e.g. Refs.~\cite{Poisson,Rohrlich,Jackson} for reviews.)

\section{Classical Position Shift}

(This and the next sections contain a summary of the relevant
results obtained in Ref.~\cite{HM4}.) In this Letter, the change in
position of the charged particle due to radiation reaction, i.e.\
the position shift, plays the central role. We present here the
position-shift formula in classical electrodynamics.  Let the
acceleration occur only for a finite time interval with $t< 0$. The
position shift, $\delta {\bf x}_{\rm C}$, that we consider is the
position at $t=0$ of the particle satisfying Eq.~(\ref{LorDir})
which would have passed through the origin at $t=0$ if $f_{\rm
LD}^\mu$ were set to zero. Let us consider a family of trajectories
${\bf X}_{\bf p}(t)$ satisfying Eq.~(\ref{LorDir}) with $f_{\rm
LD}^\mu = 0$ and the following conditions: (i) ${\bf X}_{\bf p}(0) =
0$, (ii) the momentum of the particle at $t=0$ is ${\bf p}$. Then,
the position shift can readily be found with $f^\mu_{\rm LD}$
treated as a small perturbation using the idea of ``reduction of
order"~\cite{Landau,WE} to first order in $f_{\rm LD}^\mu$:
\begin{equation}
\delta x^i_{\rm C} = - \int_{-\infty}^{0} dt\,
f_{\rm LD}^j(t)\,\frac{\partial X_{\bf p}^j(t)}{\partial
p^i}\,,  \label{first}
\end{equation}
where $f_{\rm LD}^\mu$ is evaluated on the unperturbed trajectory and
where $t$ is held fixed in the partial derivative with respect to $p^i$.
Substituting Eq.~(\ref{FLD}) we have (with $X^0_{\bf p}(t)\equiv t$)
\begin{equation}
\delta x^i_{\rm C} = e \int_{-\infty}^{0}
dt\, F_{{\rm R}\mu\nu}\frac{dX^\nu_{\bf p}}{dt}
\frac{\partial X^\mu_{\bf p}}{\partial
p^i}\,.
\end{equation}

\section{Quantum Position Shift}

In this section we present the formula for the quantum position
shift. The quantum-field-theory model corresponding to the classical
model in the previous section is the wave packet of a charged scalar
particle accelerated by an external electromagnetic field, which is
assumed to be nonzero only for a finite time in the past of the
$t=0$ hypersurface.  We assume that the external potential
$A^\mu_{\rm ext}$ (with $A^0_{\rm ext}=0$) depends only on time $t$,
and that $A^i_{\rm ext} =0$ for $t> 0$. Then the momentum of the
scalar particle is conserved (in the absence of radiation reaction).
For this reason we label the mode functions of the quantum scalar
field, chosen to have the form $e^{-ik\cdot x}$ for $t> 0$, by the
momentum ${\bf p}$. [Note, however, that the proper-velocity
$dx^i/d\tau$ of the particle is $(p^i - A^i_{\rm ext})/m$ rather
than $p^i/m$.] Let $\ket{{\bf p}}$ be the scalar one-particle state
with momentum ${\bf p}$. When the electromagnetic interaction is
turned on, to order $e^2$ this state evolves in general as
\begin{align}
\ket{{\bf p}}  \to
& [1 + i{\cal F}({\bf p})]\ket{{\bf p}} \nonumber \\
& + \frac{i}{\hbar} \int
\frac{d^3{\bf k}}{(2\pi)^3 2k} {\cal A}^\mu({\bf p},{\bf k})
\hat{a}_\mu^\dagger({\bf k})\ket{{\bf P}}\,, \label{evol}
\end{align}
with $k\equiv \|{\bf k}\|$ and
${\bf P} = {\bf p}-\hbar{\bf k}$, where the annihilation and
creation operators for the photons, $\hat{a}_\mu({\bf k})$ and
$\hat{a}_\mu^\dagger({\bf k})$, satisfy in the Feynman gauge
\begin{equation}
\left[ \hat{a}_\mu({\bf k}),\hat{a}_\nu^\dagger({\bf k}')\right]
= -g_{\mu\nu}(2\pi)^32\hbar k \delta^3({\bf k}-{\bf k}')\,.
\end{equation}
The amplitudes ${\cal F}({\bf p})$ and ${\cal A}^\mu({\bf
p},{\bf k})$ are
the forward-scattering and (one-photon-)emission amplitudes,
respectively.

The position expectation value of a wave-packet state $\ket{w}$ of the
scalar particle is defined by
\begin{equation}
\langle x^i\rangle \equiv \int d^3{\bf x}\,x^i\bra{w}\hat{J}^0\ket{w}\,,
\end{equation}
where $\hat{J}_\mu = (i/\hbar):\hspace{-0.7mm}\hat{\varphi}^\dagger
\partial_\mu \hat{\varphi} -
\partial_\mu \hat{\varphi}^\dagger \cdot \hat{\varphi}
\hspace{-0.7mm}:$ is the normal-ordered current
density operator for the charged scalar field $\hat{\varphi}$. The
$e^2$ contribution to this quantity is the quantum position shift
$\delta {\bf x}_{\rm Q}$.  This quantity is found in the limit
$\hbar\rightarrow 0$ as
\begin{align}
 \delta x^i_{\rm Q} = & - \frac{i}{2} \int \frac{d^3{\bf
k}}{2k(2\pi)^3} {\cal A}^{\mu *}({\bf p},{\bf k})
\stackrel{\leftrightarrow}{\partial}_{{p}^i} {\cal A}_\mu( {\bf
p},{\bf k}) \nol & - \hbar \partial_{p^i} {\rm Re}\,{\cal F}({\bf p})\,.
\label{xiQ}
\end{align}
In Ref.~\cite{HM4} the right-hand side was shown to be
equal to the classical position shift, $\delta x^i_{\rm C}$,
given by Eq.~(\ref{first}) with
$f_{\rm LD}^\mu$ found in Eq.~(\ref{FLD2}).  This
calculation did not give much mathematical
insight into why the quantum position
shift in the $\hbar \to 0$ limit
is identical to the classical one.
In this Letter we express
the quantum position shift given by Eq.~(\ref{xiQ}) in terms of
Green's functions and clarify the reason why it
agrees with the classical expression.
In the next two sections
we shall analyze the contribution from the emission amplitude and that
from the forward scattering amplitude in Eq.~(\ref{xiQ}) in turn.

\section{Contribution from the Emission Amplitude}

The emission contribution to the quantum position shift (\ref{xiQ})
is
\begin{equation}
\label{deltaem}
\delta x_{\rm em}^i = -\f{i}{2}\int \f{d^3k}{2k(2\pi)^3}
{\cal A}^{\mu
*}({\bf p},{\bf k})
\stackrel{\leftrightarrow}{\partial}_{p^i} {\cal A}_\mu ({\bf p},{\bf
k})\,.
\end{equation}
Using the WKB scalar mode
functions in the presence of external field, one finds that the emission
amplitude coincides with that from a classical point
charge to order $\hbar^0$~\footnote{This is the case if the external
electromagnetic potential depends only on one time {\em or space}
coordinate~\cite{HM4}.}.  That is~\cite{HM4},
\begin{equation}
{\cal A}^\mu({\bf p},{\bf k}) =
- \int_{-\infty}^{+\infty} e^{ik\cdot x} j^\mu(x)\,,
\end{equation}
where the current $j^\mu(x)$ is given by
\begin{equation}
j^\mu(x) = \frac{dx^\mu}{dt} \delta^3({\bf x}-{\bf X}_{\bf p}(t))\chi(t)
\,.  \label{fakecurrent}
\end{equation}
Here the function $\chi(t)$ is a cut-off factor that takes the
value $1$ when the external force is nonzero and smoothly becomes zero
for large $|t|$.
The classical field emitted from the current (\ref{fakecurrent}) is
\begin{equation}
A^\mu_{-}(x) = \int d^4x' G^{\ph\mu}_{-\ph\nu'}(x-x')
j^{\nu'}(x')\,. \label{retarded}
\end{equation}
Now, the retarded Green's function can be written as~\cite{Itzykson}
\begin{multline}
G_{-\mu\nu'}(x-x') = i g_{\mu\nu'}\theta(t-t')\int \frac{d^3{\bf
k}}{2k(2\pi)^3}\\
\times \left[ e^{-ik\cdot (x-x')} - e^{ik\cdot(x-x')} \right]\,.
\end{multline}
These two equations and Eq.~(\ref{fakecurrent}) imply that
\begin{equation}
A^\mu_{-}(x) =
-i \int\f{d^3k}{2k(2\pi)^3}
\[ {\cal A}^\mu ({\bf p},{\bf k})e^{-ik\cdot x} -
{\cal A}^{\mu *}({\bf p},{\bf k}) e^{ik\cdot x} \]\,,
\end{equation}
for large enough $t$ such that $\chi(t)=0$.
Then, by a straightforward calculation one finds that the emission
contribution to the position shift given by Eq.~(\ref{deltaem}) can be
expressed as follows:
\beq
\label{fieldposshift} \delta x_{\rm em}^i = -\f{1}{2} \int_{t=T} d^3{\bf x}
\(\partial_{p^i}A^\mu_{-}\)\stackrel{\leftrightarrow}{\partial}_t
A_{-\mu}\,,
\eeq
where $t=T$ is far into the future so that $\chi(T)=0$.
By substituting Eq.~(\ref{retarded}) and using
\begin{multline}
\int_{t=T} d^3{\bf x} \,
{{G_{-}}^\mu}_{\alpha''}(x-x'')\stackrel{\leftrightarrow}{\partial}_t
G_{-\mu\beta'}(x-x') \\
= - 2G_{{\rm R}\alpha''\beta'}(x''-x')\,,
\end{multline}
for $x_0 > {\rm max}(x^{\prime}_0,x^{\prime\prime}_0)$, we obtain
\begin{align}
\delta x_{\rm em} & = \int d^4x d^4x'\,
\partial_{p^i} j_\mu(x) G_R^{\mu\nu'}(x-x')j_{\nu'}(x')\nonumber \\
  & = \int
d^4x\, \partial_{p^i}j^\mu(x)A_{{\rm R}\mu}(x)\,, \label{standard0}
\end{align}
where the regular field $A^\mu_{\rm R}(x)$ is defined by
Eq.~(\ref{regularfield}).  By a
calculation analogous to the derivation of
the Lorentz force from the standard Lagrangian for a point charge in
an external electromagnetic field
(see, e.g. Ref.~\cite{Jackson}), we find from Eq.~(\ref{standard0})
\begin{align}
\delta x_{\rm em}^i  & = e\int dt F_{{\rm R}\mu\nu}\frac{dX^\nu_{\rm
p}}{dt}
\left( \frac{\partial X^\mu_{\bf p}}{\partial p^i}\right)_t \nonumber \\
& =  -e\int_{-\infty}^0 dt\,f^j_{\rm LD}\,\left(
\frac{\partial X^j_{\bf p}}{\partial p^i}\right)_t \,.
\end{align}
Surface terms can be dropped thanks to the cut-off function
$\chi(t)$.  The final result is independent of the choice of $\chi(t)$.
We have made the upper bound of the $t$-integration to $t=0$ in the last
line because $f^j_{\rm LD} =0$ for $t > 0$.  Thus, we have shown that
the contribution from the emission of a photon to the position shift
agrees with the classical counterpart using the Green's function method.

\section{Contribution from the Forward-Scattering Amplitude}

We now turn to the forward-scattering contribution to the
position shift given by
\beq
\delta x_{\rm for}^i =  - \hbar\partial_{p^i} {\rm
Re} \,{\cal F}({\bf p})\,.
\eeq
This contribution vanishes in the end~\cite{HM4}.
More precisely,
the leading order terms of the real part of the forward-scattering
amplitude
are exactly cancelled by the contribution from the mass counter-term,
i.e.\ it is eliminated to order $\hbar^0$
by the mass renormalization.   Here we shall see that the field
generated by the singular Green's function appears in the calculation of
$\delta x^i_{\rm for}$.

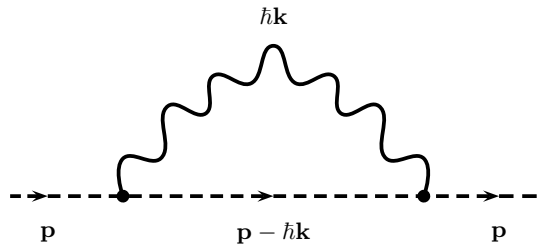
\begin{figure}
\begin{center}
\begin{pspicture}(-4,-1)(4,2.5)
\psline[linewidth=0.5mm,linestyle=dashed]{->}(-3.5,0)(-3,0)
\psline[linewidth=0.5mm,linestyle=dashed]{->}(-3,0)(0,0)
\psline[linewidth=0.5mm,linestyle=dashed]{->}(0,0)(3,0)
\psline[linewidth=0.5mm,linestyle=dashed](3,0)(3.5,0)
\pscurve[linewidth=0.5mm]{*-*}(-2,0)(-2,.5)(-1.5,.5)%
(-1.4,1.2)(-.9,1.1)(-.8,1.6)(-.3,1.5)(0,2)(.3,1.5)(.8,1.6)(.9,1.1)%
(1.4,1.2)(1.5,.5)(2,.5)(2,0)
\rput(-3,-.5){${\bf p}$}
\rput(0,-.5){${\bf p}-\hbar{\bf k}$}
\rput(3,-.5){${\bf p}$}
\rput(0,2.4){$\hbar{\bf k}$}
\end{pspicture}
\caption{The one-loop diagram contributing to the
forward-scattering amplitude: the dashed and wavy lines represent
the scalar and photon propagators, respectively.}
\label{feynman}
\end{center}
\end{figure}

The forward-scattering amplitude comes from the one-loop diagram shown
in Fig.~\ref{feynman} (and the tadpole diagram, which can be ignored in
the end). For the contribution from the intermediate {\em particle}
state (as opposed to {\em anti-particle} state), we
divide the momentum integral for the virtual photon in this loop diagram
into two parts; one with momentum
$\hbar\|{\bf k}\|$ less than $\hbar^{\alpha}\lambda$ and the other
with momentum larger than $\hbar^{\alpha}\lambda$, where $\alpha$ and
$\lambda$ are constants.  We choose $\alpha$ to satisfy
$\frac{3}{4} < \alpha < 1$ as in Ref.~\cite{HM4}, although the
condition $\alpha <1$ will suffice for the purpose of this Letter.
Denoting the first part with the virtual-photon momentum below the
cut-off by ${\cal F}^{<}({\bf p})$, we find to lowest order in
$\hbar$~\cite{HM4}
\begin{align}
\hbar {\cal F}^{<}({\bf p}) = & - ie^2 \int_{k \leq
\hbar^{\alpha-1}\lambda}
\frac{d^3{\bf k}}{(2\pi)^32k}\int_{-\infty}^{+\infty}dt
\int_{-\infty}^{+\infty}dt'
\nol &\times \theta(t-t')\frac{dX_{\bf p}^\mu}{dt}
\frac{dX_{{\bf p}\mu}}{dt'}
e^{ik(t'-t) - i{\bf k}\cdot({\bf X}_{\bf p}(t')-{\bf X}_{\bf p}(t))}\,.
\end{align}
Here we have corrected some misprints in the corresponding equation,
Eq.~(A37), in Ref.~\cite{HM4}.
In the classical limit $\hbar\to 0$, the ${\bf k}$-integration will have
no restriction because
$\hbar^{\alpha-1}\lambda \to \infty$.  Recalling that the Feynman
propagator is given by
\begin{align}
G_F^{\mu\nu'}(x-x') = & -i\hbar g^{\mu\nu'}
\int \f{d^3k}{(2\pi)^3 2k}\nonumber  \\
 & \times \[ \theta(x_0-x_0')e^{-ik\cdot(x-x')} +
(x\leftrightarrow x') \]\,,
\end{align}
we find in this limit
\begin{equation}
\hbar{\cal F}^{<}({\bf p})
= \frac{1}{2\hbar}\int d^4 x d^4
x'\,j_\mu(x)j_{\nu'}(x')G_F^{\mu\nu'}(x-x')\,.
\end{equation}
Let us write Feynman propagator as the sum of
the real and imaginary parts:
\begin{align}
G_F^{\mu\nu'}(x-x') &=
-\hbar G_S^{\mu\nu'}(x-x') - i\hbar G^{(1)\mu\nu'}(x-x')\,,
\end{align}
where
\begin{equation}
\hbar G^{(1)\mu\nu'}(x-x') =
\bra{0} \left\{ \hat{A}^\mu(x),\hat{A}^{\nu'}(x') \right\} \ket{0}\,,
\end{equation}
with $\hat{A}^\mu(x)$ being the quantum electromagnetic potential.
The imaginary part ${\rm Im}\,{\cal F}^{<}({\bf p})$
equals half the emission probability as required by unitarity:
\begin{equation}
{\rm Im} \, {\cal F}^{<}({\bf p})
= -\f{1}{2\hbar} \int \f{d^3 {\bf k}}
{2k(2\pi)^3} {\cal A}_\mu^*({\bf p},{\bf
k}){\cal A}^\mu({\bf p},{\bf k})\,.
\end{equation}
(It turns out that the contribution to the
forward-scattering amplitude above the cut-off has no imaginary part
in the $\hbar \to 0$ limit~\cite{HM4}
).
Using the symmetry of $G_S^{\mu\nu'}(x-x')$, we obtain
\begin{equation}
-\hbar \partial_{p^i} {\rm Re} \, {\cal F}^{<}({\bf p}) = \int d^4x\,
 \partial_{p^i} j^\mu (x) A_{S\mu}(x)\,, \label{singcont}
\end{equation}
where $A_S^\mu(x)$ is the singular part of the self-field given by
\begin{equation}
A_S^\mu(x) = \int d^4 x'\,G_S^{\mu\nu'}(x-x')j_{\nu'}(x')\,.
\end{equation}
If we add Eq.~(\ref{singcont})
to the emission contribution, i.e.\
Eq.~(\ref{standard0}), then
\begin{equation}
\delta x_{\rm em} - \hbar\partial_{p^i}{\rm Re}\,{\cal F}^{<}({\bf p})
= \int d^4x\,
 \partial_{p^i} j_\mu (x)A_{-}^\mu(x)\,,
\end{equation}
where the self-field, $A_{-}^\mu(x)$, is given by Eq.~(\ref{retarded}).
Thus, the one-photon emission process and the low-energy part of the
forward-scattering process are incorporated
in the classical self-field $A_{-}^\mu(x)$ if one sees this field
from the
viewpoint of quantum derivation of the self-force. The
regular part, $A_R^\mu(x)$, of the self-field in classical
electrodynamics corresponds to the emission process in QED
and the singular part, $A_S^\mu(x)$,
to the low-energy forward-scattering
process.  The remaining high-energy and intermediate anti-particle state
contributions to the
forward-scattering amplitude in QED
have no classical counterpart. The forward-scattering
contribution as a whole vanishes if one includes
the quantum mass counter-term, as was shown in Ref.~\cite{HM4}.

\section{Summary}

In this Letter we re-expressed the $\hbar\to 0$ limit
of the position shift due to radiation reaction in QED, Eq.~(\ref{xiQ}),
in terms of the Green's functions for the electromagnetic field, thus
making the relation between the classical and quantum derivations of the
radiation-reaction force more transparent.  We believe that the insight
gained in this Letter will be useful in finding the electomagnetic and
gravitational radiation-reaction
forces in quantum field theory in curved spacetime, which are to be
compared with results
in classical field theory in the literature~\cite{Sasaki,Wald,Quinn}.

\end{document}